\documentclass{PoS}
\usepackage{slashed}
\usepackage{ dsfont }
\usepackage{amsmath}
\usepackage{amsthm} 
\usepackage{amssymb}
\usepackage{amsfonts}

\title{Dark Matter phenomenology of intersecting D6-branes with a St\"uckelberg portal}

\ShortTitle{Dark Matter phenomenology of intersecting D6-branes with a St\"uckelberg portal}

\author{\speaker{Victor Martin Lozano}%
         \thanks{Talk based on Ref.~\cite{1}, done in collaboration with Miguel Peir\'o and Pablo Soler.}\\
              Instituto de F\'{\i}sica Te\'{o}rica
      UAM/CSIC \& Departamento de F\'{\i}sica Te\'{o}rica, \\Universidad Aut\'{o}noma de Madrid, E-28049
      Madrid, Spain\\
        E-mail: \email{victor.martinlozano@uam.es}}

\abstract{We discuss the possibility that a St\"uckelberg portal connects both Standard Model and Dark Matter sectors. The particle responsible of this portal is the lightest $Z'$ boson that induces isospin-violating interactions. This property leads to a rich phenomenology for the direct detection and collider experiments that can constraint the parameter space of this kind of models and can be tested in the future.\\
\begin{flushright}
FTUAM-15-33 \\
IFT-UAM/CSIC-15-109 
\end{flushright}}

\FullConference{18th International Conference From the Planck Scale to the Electroweak Scale \\
                 25-29 May 2015\\
                 Ioannina, Greece }

\begin{document}

\section{Introduction}

The non-gravitationally interactions between the Standard Model (SM) particles and the Dark Matter (DM) particles is nowadays a key point in Physics since from that we can infer the nature of the dark sector of the Universe.

One simple idea is to locate the SM content and the DM content in different sectors so the only possible interaction is the gravitational one. However for different reasons such as the existence of a DM density of the Universe we know that there should be a connection between both sectors. This interaction can be achieved through the idea of portals \cite{2,3,4,5,6,7,8,9,10,11,12,13,14,15,16}. The definition of a portal is simple, a particle or a set of particles that could live in two different sectors connect them through the interactions with the particles of each sector. The nature of the mediator could be diverse, among all the possibilities the most important ones are the Higgs portal \cite{17,18} and the $Z'$ portal\cite{19,20,21,22,23,24,25,26}. The former one has been studied widely in the literature relating it with the new scalar boson found in the LHC. The latter is also well-studied in the literature and we will focus on this kind of portals due to the possibility of connect it to an ultraviolet complete theory.
\vspace*{5.0mm}

\begin{figure}
  \centering
\includegraphics[scale=0.55]{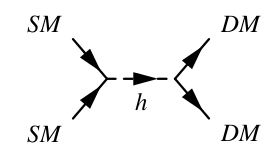}\hspace{2cm}
\includegraphics[scale=0.55]{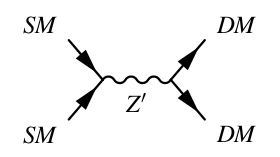}
\caption{Possible diagrams that connect the SM sector with the DM one for Higgs portal (left) and $Z'$ portal (right).}
\label{fig:hz}
\end{figure}

The $Z'$ portal can be built as the SM sector with a number $n$ of $U(1)_V$ gauge bosons and the dark sector as a semi-simple lie group, $G_h$ with number $m$ of $U(1)_V$ gauge bosons,
\begin{eqnarray}
\,\,\,SU(3)_{c}\,\times\, SU(2)_{L}\,\times\,U(1)_{v}^n~\times~U(1)_{h}^m\,\times\, G_{h}.  \nonumber\\[-10pt]
\!\!\!\underbrace{ \hphantom{SU(3)_{c}\,\times\, SU(2)_{L}\,\times\,U(1)_{v}^n}}_{SM} \hphantom{~\times~}\underbrace{ \hphantom{U(1)_{h}^m\,\times\,G_{h}}}_{DM}\nonumber
\end{eqnarray}
In general there are kinetic or mass mixing terms between the gauge bosons of both sectors ($A^n_v, A^m_h$) that can originate the connection between the particles of the SM and the DM particles. The origin of the mass terms is the St\"uckelberg mechanism that provide masses to the $U(1)$ gauge bosons through a mass matrix with relatively large non-diagonal entries assuring the mixing. This kind of models with St\"uckelberg mechanisms are easily embedded into string theory models \cite{8,9}. Once the Lagrangian is diagonalized the lightest $Z'$ is the one that plays a r\^ole in the phenomenology since the others are heavy to have an effect on that. The way this lightest $Z'$ boson couples to the SM particles is determined by the diagonalization of the mass matrix and define the interaction between the DM particles to the SM particles. The most important thing to note here is the fact that in general the $Z'$ boson couples differently to left and right particles inducing a source of isospin violation in the connection with the DM. This property affects directly to the phenomenology of the DM particle and also to the searches of it. For example, for the direct detection searches an isospin-violating DM could evade the limits imposed by the experiments and they have to be computed simulating the number of events. The DM could be produced in colliders where the isosping-violating nature also arises. So in this work we studied the phenomenology of the isospin-violating DM and its effects in the direct detection searches and the collider searches as well, showing that both experimental methods are complementaries to disentangle the nature of DM.

\section{Effective Lagrangian and $Z'$ eigenstates}

The St\"uckelberg portal consists of the mixing of massive $U(1)$ gauge bosons from different sectors. The general construction in terms of an effective field theory can be described in terms of the following Lagrangian \cite{8,9},
\begin{equation}
\mathcal{L}=-\frac{1}{4}\vec{F}^T\cdot f \cdot \vec{F} - \frac{1}{2} \vec{A}^T\cdot M^2\cdot \vec{A} + \sum_{\alpha}\bar{\psi}_\alpha\left(i\slashed{\partial} + \vec{Q}^T_\alpha \cdot \slashed{\vec{A}}\right)\psi_\alpha,
\end{equation}
where $\vec{F}=d\vec{A}$ is the field strength of the $U(1)$ gauge bosons encoded in the vector $\vec{A}^T=(A_1 ... A_{n+m})$. The matrices $f$ and $M$ are the kinetic and mass matrices, and in general they are non-diagonal. However for the sake of simplicity we only consider $M$ to be non-diagonal while keeping $f$ to be diagonal~\footnote{In general the non-diagonal entries of $f$ are originated through loop processes that are suppressed and can be neglected safely.}.  For a given matter field, $\psi_\alpha$, the charge vectors $\vec{Q}_\alpha$ will have non-zero entries either for the visible sector or the hidden one.

In order to acquire masses for the Abelian gauge bosons $\vec{A}$ are coupled to a set of axionic periodic fields $\phi^i\sim \phi^i +2\pi$ in the following manner in the Lagrangian,
\begin{equation}
\mathcal{L}_M=-\frac{1}{2}G_{ij}(\partial \phi^i-k_a^iA^a)(\partial \phi^j-k_b^jA^b),
\label{eq:massstu}
\end{equation}
where $G_{ij}$ is the metric in the space of $\phi^i$ fields and it is a positive-definite matrix. The factors $k_a^i$ are the axionic charges and under the appropriate normalization they can be assumed to be integers, $k_a^i\subset \mathds{Z}$. After gauge fixing the $U(1)$ symmetries, the Abelian gauge bosons absorb the axions and acquire a mass through the St\"uckelberg mechanism. The corresponding mass matrix from Eq.~\ref{eq:massstu} reads,
\begin{equation}
M^2=K^T\cdot G\cdot K.
\end{equation}
Here $K$ is the axionic-charge matrix that is, in general, non-diagonal making the mass matrix $M$ also non-diagonal. So, the fact that $M$ has large non-diagonal entries implies a non-negligible mixing between the visible and the hidden sectors.

To study the physical implications of this mixing it is convenient to work on the mass basis where the kinetic terms are canonical and the mass matrix is diagonal. The canonical kinetic term could be obtained with a linear transformation,
\begin{equation}
\vec{A}\equiv \Lambda \cdot\vec{A}', \quad {\rm with} \quad \Lambda^T\cdot f\cdot\Lambda =1.
\end{equation}
If there is no kinetic mixing, $f={\rm diag}(g_1^{-2},...,g_N^{-2})$, the transformation is simplified, $\Lambda=\rm{diag}(g_1,...,g_N)$. After such transformation the new Lagrangian can be written
\begin{equation}
\mathcal{L}= -\frac{1}{4}\vec{F}'^2-\frac{1}{2}\vec{A}'^T\cdot \Lambda^T\cdot M^2\cdot \Lambda\cdot \vec{A}' + \sum_{\alpha}\bar{\psi}_\alpha\left(i\slashed{\partial} + \vec{Q}^T_\alpha \cdot \Lambda \cdot \slashed{\vec{A}}'\right)\psi_\alpha.
\end{equation}
In order to diagonalise the mass matrix an orthogonal transformation $\mathcal{O}$ must be performed so
\begin{equation}
\tilde{M}^2\cdot \vec{v}_i =m_i^2\vec{v}_i\quad \Rightarrow\quad \mathcal{O}=(\vec{v}_1 ...\vec{v}_N),
\end{equation}
where $\tilde{M}=\Lambda^T\cdot M^2\cdot \Lambda$. If we define $\vec{v}_i'\equiv \Lambda \cdot \vec{v}_i$ and $\vec{A}'\equiv \mathcal{O}\cdot \vec{A}''$ the physical Lagrangian reads,
\begin{equation}
\mathcal{L}=-\frac{1}{4}\vec{F}_i''^2- \frac{1}{2}m_i^2\vec{A}_i''^2+\sum_\alpha \bar{\psi}_\alpha(i\slashed{\partial}+\vec{g'}_{\alpha}^T\cdot\slashed{\vec{A}}'')\psi_\alpha.
\end{equation}
Now the coupling of a matter field $\psi_\alpha$ to a vector boson $A_i''$ is given by
\begin{equation}
g'_{\alpha} {}^{(i)}=\vec{Q}_\alpha^T\cdot \vec{v}'_{(i)}.
\label{eq:couplingZ}
\end{equation}
This means that now the couplings to the vector bosons are combinations of the original charges. It is important to remark that for massless eigenvectors the corresponding gauge bosons are massless and have quantized charges, so choosing an appropriate form of the $K$ matrix these gauge bosons will couple only to one sector of the theory.

It is important to remark that the St\"uckelberg portal in the way it is constructed as an effective field theory can be embedded naturally in the frame of string theory. Type IIA string theory with intersecting D6-branes provides both sectors and the couplings of the gauge bosons to axions in such a way that the St\"uckelberg mechanism is realised in a natural way \cite{8,9}.

\begin{figure}
\begin{center}
\includegraphics[scale=0.25]{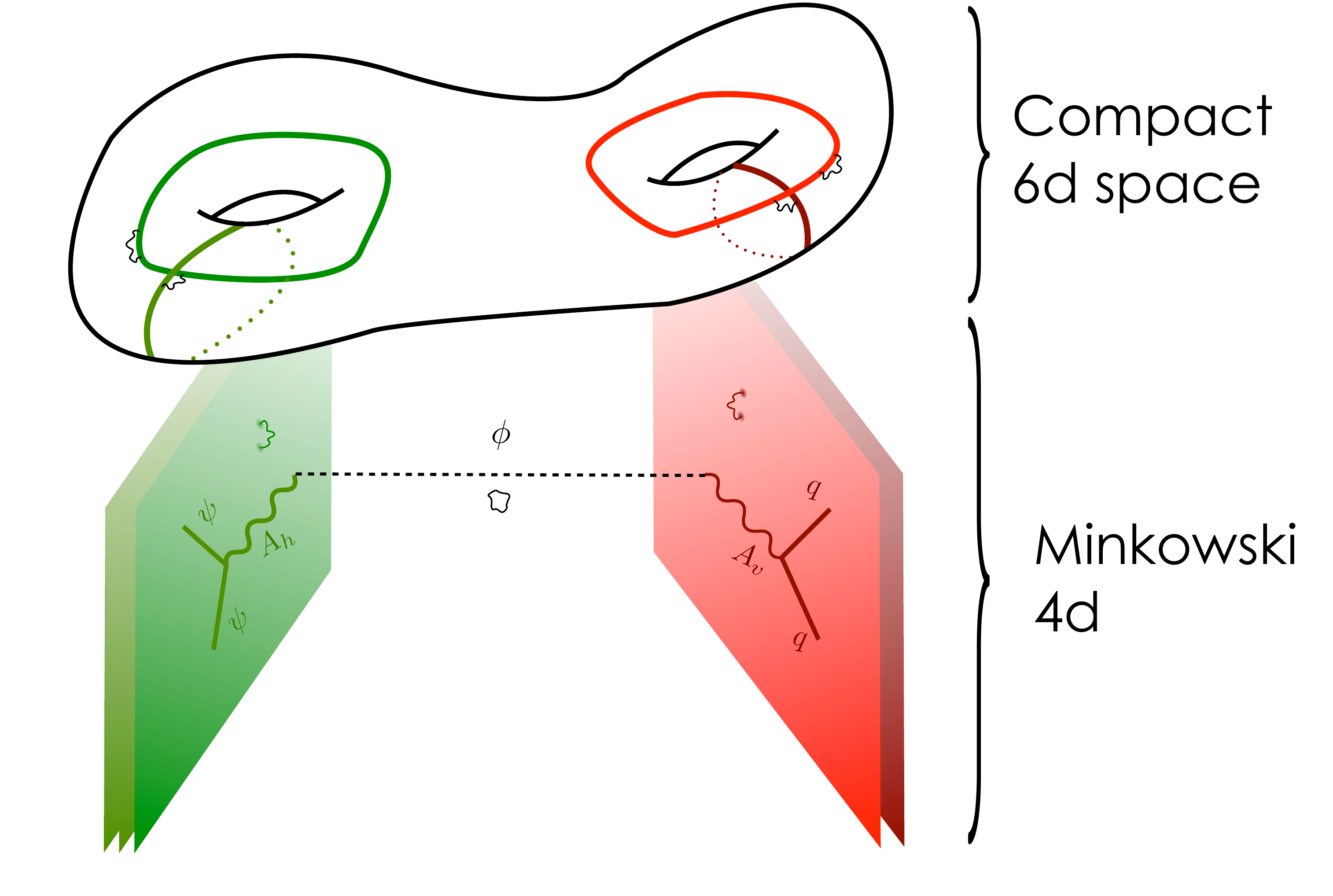}\hspace{2cm}
\includegraphics[scale=0.55]{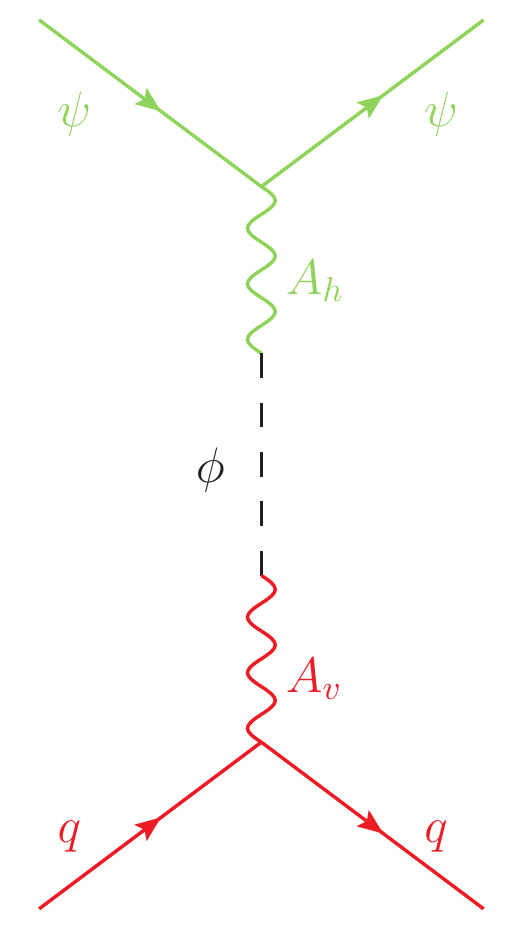}
\label{fig:branes}
\caption{Left: schematic representation of a hidden sector scenario. Red branes represent the visible sector while green ones are from the hidden sector. Right: diagram showing the interaction of both sectors through the mixing of different string axions, $\phi$, and the vector gauge bosons $A_v$ and $A_h$.}
\end{center}
\end{figure}
\subsection{SM fermion couplings to $Z'$}

Due to the fact that the St\"uckelberg portal can be embedded into string models we focus on the \emph{Madrid models} \cite{27} as it was done in Refs.\cite{8,9,1}. The idea is to introduce four stacks of D6-branes to get $U(3)_A\times U(2)_B \times U(1)_C \times U(1)_D$ gauge group as the visible group. In Table \ref{tab:coup} the charges of the SM fermions under the previous gauge group are depicted. With this assignment the hypercharge corresponds to the linear combination
\begin{equation}
Q^Y=\frac{1}{6}(Q_A-3Q_C+3Q_D),
\end{equation}
that corresponds to the only massless gauge boson and only the visible particles are coupled to it. The coupling to the visible matter field to the hypercharge vector boson is
\begin{equation}
g^Y_\alpha=e Q^Y_\alpha=\frac{e}{6}(Q_{\alpha A}-3Q_{\alpha C}+3Q_{\alpha D}).
\end{equation}
In general the couplings of the lightest $Z'$ boson to the matter fields $\psi_\alpha$ is written as
\begin{equation}
g^{Z'}_\alpha=aQ_{\alpha A}+ bQ_{\alpha B} + cQ_{\alpha C} + dQ_{\alpha D} + \sum_{i=1}^m h_i Q_{\alpha i}^{h},
\end{equation}
where the sum over $1, ... , m$ corresponds to the hidden sector. The parameters $a$, $b$, $c$, $d$ and $h_i$ are precisely the entries of the vector $\vec{v}'_{Z'}=(a,b,c,d;h_1 ... , h_m)$ of Eq.~(\ref{eq:couplingZ}).

Using the values of Table~\ref{tab:coup} we can get the couplings of the SM particles to the lightest $Z'$ that are,
\begin{eqnarray}
C_u^V=(b+c),& C_u^A=(2a+b-c),\\
C_d^V=(b-c),& C_d^A=(2a+b+c),\\
C_t^V=(-b-c),& C_t^A=(2a-b-c),\\
C_b^V=(-b+c),& C_b^A=(2a-b+c),\\
C_\ell^V=(-b-c),& C_\ell^A=(-b+c-2d).
\end{eqnarray}
As we can see the vectorial and axial couplings to the same particles are different originating isospin-violating effects that as we can see in the next section affect directly to the phenomenology of this model.
\begin{table}
\begin{center}
\begin{tabular}{|c|c|c|c|c|c|}
\hline
Matter field & $Q_A$ & $Q_B$ & $Q_C$ & $Q_D$ & $Y$\\
\hline
$Q_L$&1&-1&0&0&1/6\\
$q_L$&1&1&0&0&1/6\\
$U_R$&-1&0&1&0&-2/3\\
$D_R$&-1&0&-1&0&1/3\\
\hline
$L$&0&-1&0&-1&-1/2\\
$E_R$&0&0&-1&1&1\\
$N_R$&0&0&1&1&0\\
\hline
\end{tabular}
\caption{SM spectrum and $U(1)_i$, charges in the four stack models of Ref.\cite{27}. Anomaly cancellation requires the three quark families to be divided into two $Q_L$ doublets and two antidoublets $q_L$.}
\label{tab:coup}
\end{center}
\end{table}

\section{Isospin violation from the St\"uckelberg mechanism in light of LHC and LUX results}

\begin{figure}
\begin{center}
\includegraphics[scale=0.35]{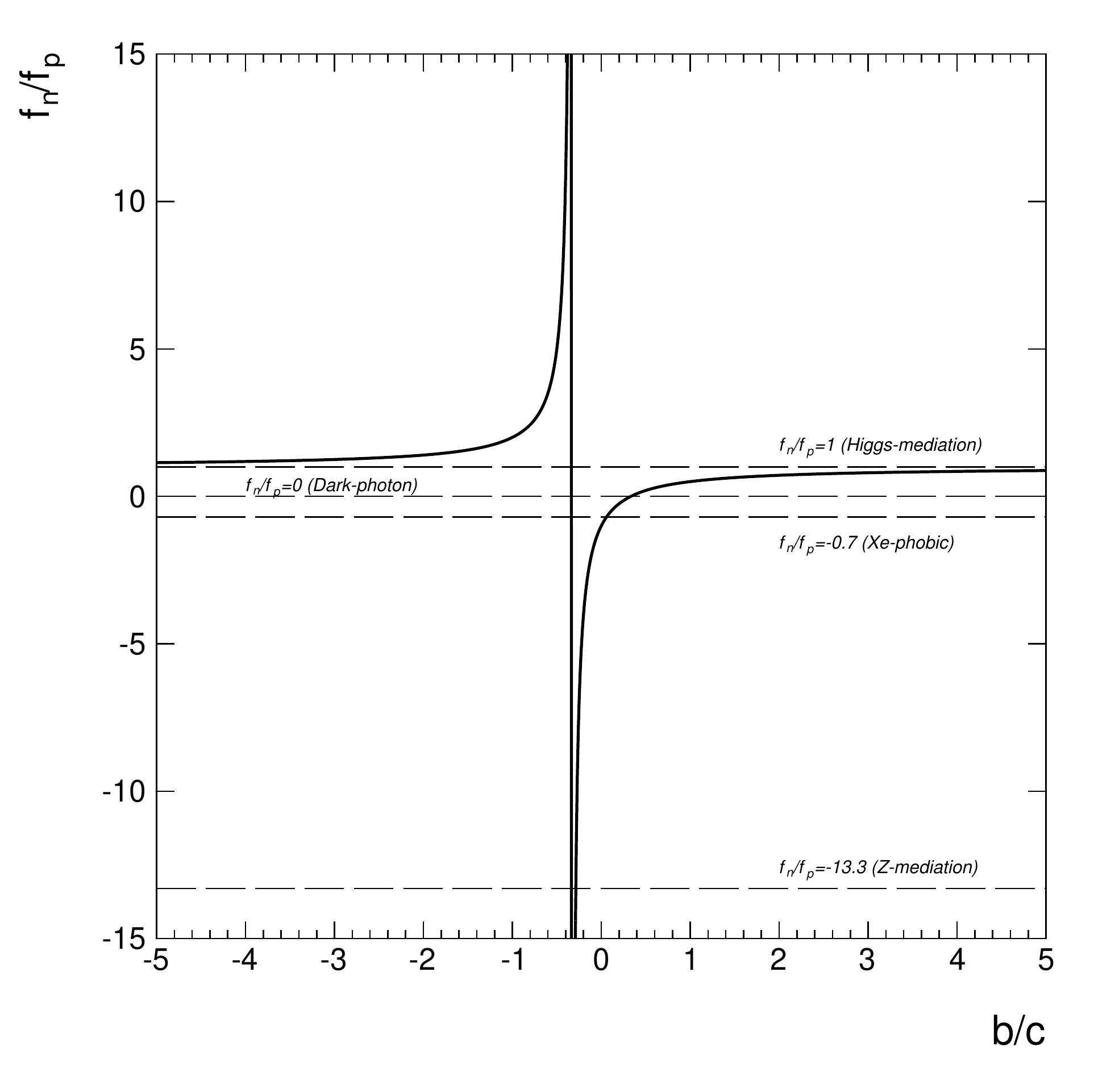}\hspace{1cm}
\includegraphics[scale=0.35]{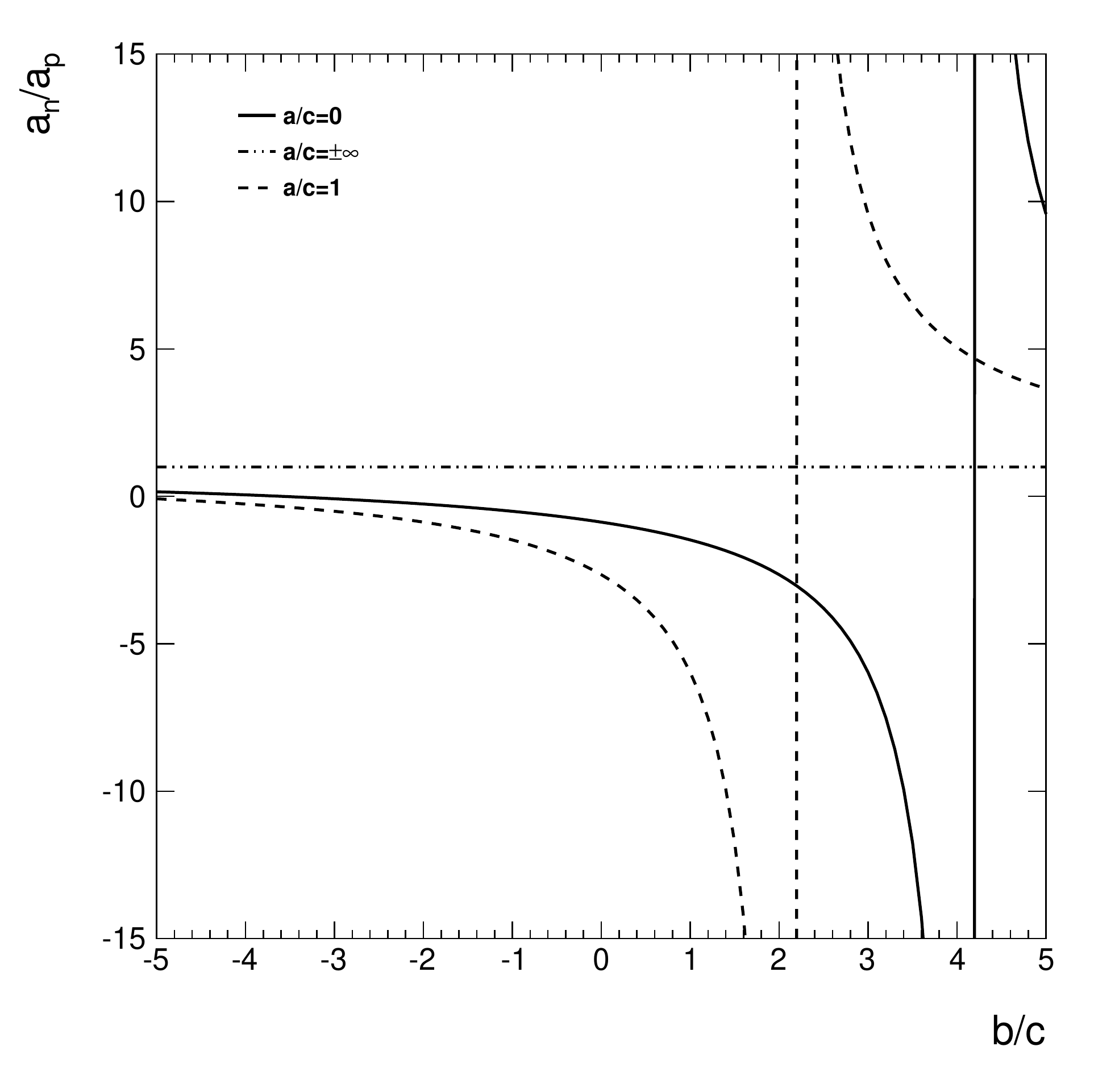}
\label{fig:fnfp}
\caption{Left: amount of isospin violation for SI interactions, $f_n/f_p$, as a function of $b/c$. Some representative values of $f_n/f_p$ are shown as horizontal dashed lines. Right: ratio between the coupling of DM to neutrons and protons, $a_n/a_p$, for the SD interactions as a function of $b/c$. For $a/c$ we have taken different limits, $a/c\to 0$ (solid line), $a/c\to \pm \infty$ (dot-dashed line), and $a/c=1$ (dashed line).}
\end{center}
\end{figure}

As we have seen in the previous section the $Z'$ couples differently to up and down quarks. In light of direct detection experiments this property is important \cite{28} since for every nucleus we have different number of protons than neutrons, \emph{i.e.} different number of up and down quarks. That means that different nucleai have different sensitivities to the scattering off dark matter. The effective Lagrangian of the spin-independent (SI) interaction between the DM particle and the protons and the neutrons is,
\begin{equation}
\mathcal{L}_{SI}=f_p(\bar{\psi}\gamma_\mu\psi)(\bar{p}\gamma^\mu p) + f_n(\bar{\psi}\gamma_\mu \psi)(\bar{n}\gamma^\mu n),
\end{equation}
where the couplings of the DM to the protons ($f_p$) and neutrons ($f_n$) are given by \cite{29}
\begin{equation}
f_p=2b_u+b_d=\frac{h}{2m_{Z'}^2}(3b+c),\quad fn=2b_d+b_u=\frac{h}{2m_{Z'}^2}(3b-c).
\end{equation}
So the amount of isospin violation can be obtained with the ratio of these two quantities,
\begin{equation}
\frac{f_n}{f_p}=\frac{3b/c-1}{3b/c+1}.
\end{equation}
We see that it only depends on two of the variables of the model, namely $b$ and $c$. The value of $f_n/f_p$ determines the sensitivity of a nucleus to the scattering. In Fig.\ref{fig:fnfp} we show the dependence of $f_n/f_p$ as a function of $b/c$, where it is easy to see that all possible values can be achieved. For the spin-dependent (SD) interaction is similar, the Lagrangian could be read as
\begin{equation}
\mathcal{L}_{SD}=a_p(\bar{\psi}\gamma_\mu\gamma_5\psi)(\bar{p}\gamma^\mu\gamma_5 p) + a_n(\bar{\psi}\gamma_\mu\gamma_5 \psi)(\bar{n}\gamma^\mu\gamma_5 n),
\end{equation}
where $a_n$ and $a_p$ are similar to the corresponding values for the spin-independent case but for the axial-vector coupling. The ratio between these two quantities becomes more complicated due to the fact that the appearance of the quantities $\Delta_q^{p(n)}$ that relate the nucleon to the quarks. The ratio can be written as
\begin{equation}
\frac{a_n}{a_p}\frac{\Delta_u^n + \frac{2a/c+b/c+1}{2a/c+b/c-1}(\Delta_d^n +\Delta_s^n)}{\Delta_u^p +\frac{2a/c+b/c+1}{2a/c+b/c-1}(\Delta_d^p + \Delta_s^p)},
\end{equation}
we can see its behaviour in Fig.\ref{fig:fnfp} as a function of $b/c$ for given values of $a/c$. 

The fact that we have isospin violation makes us to evaluate the bounds from the direct detection experiments and colliders taking into account this property.

\begin{figure}
\begin{center}
\includegraphics[scale=0.35]{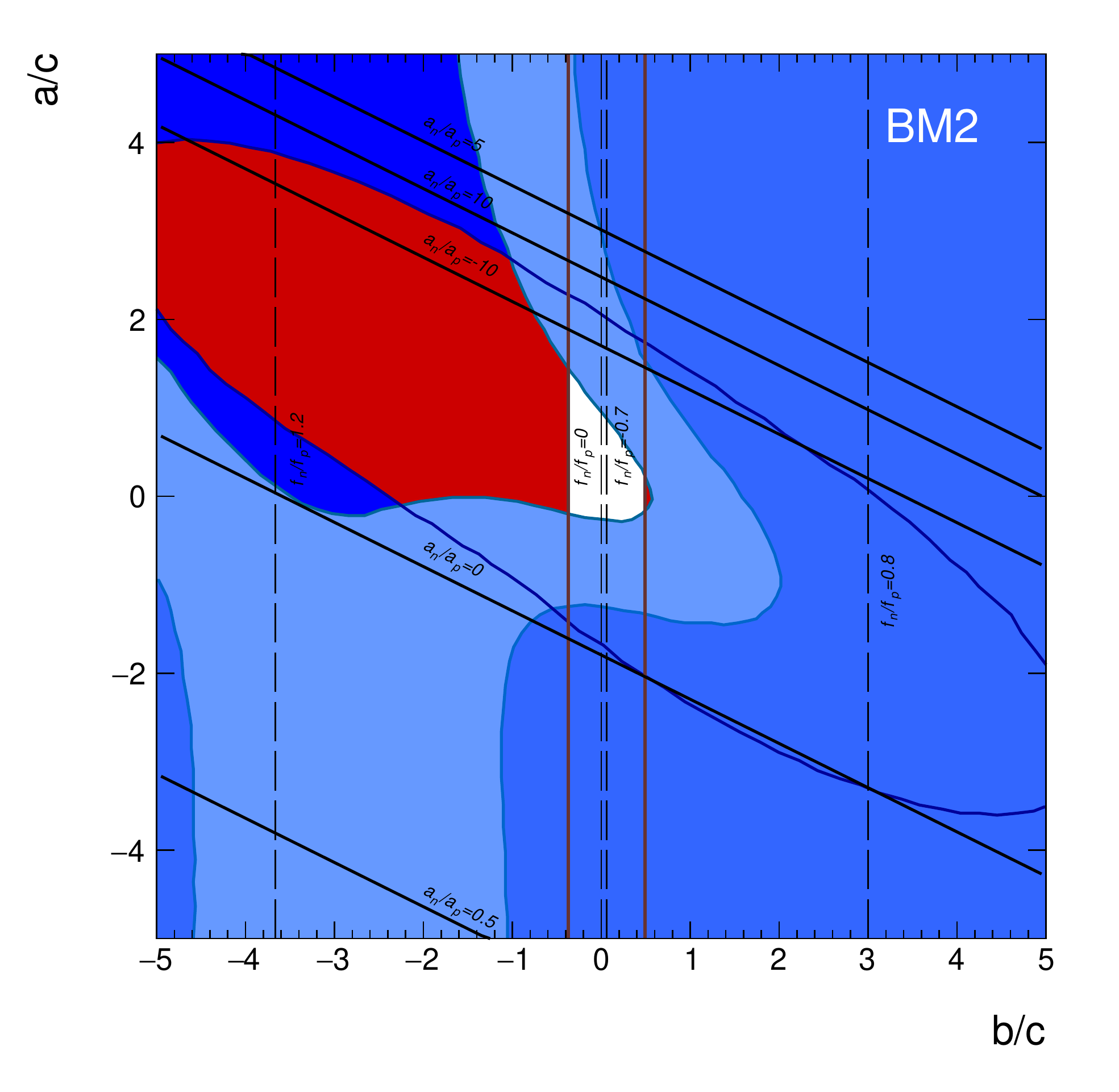}\hspace{1cm}
\includegraphics[scale=0.35]{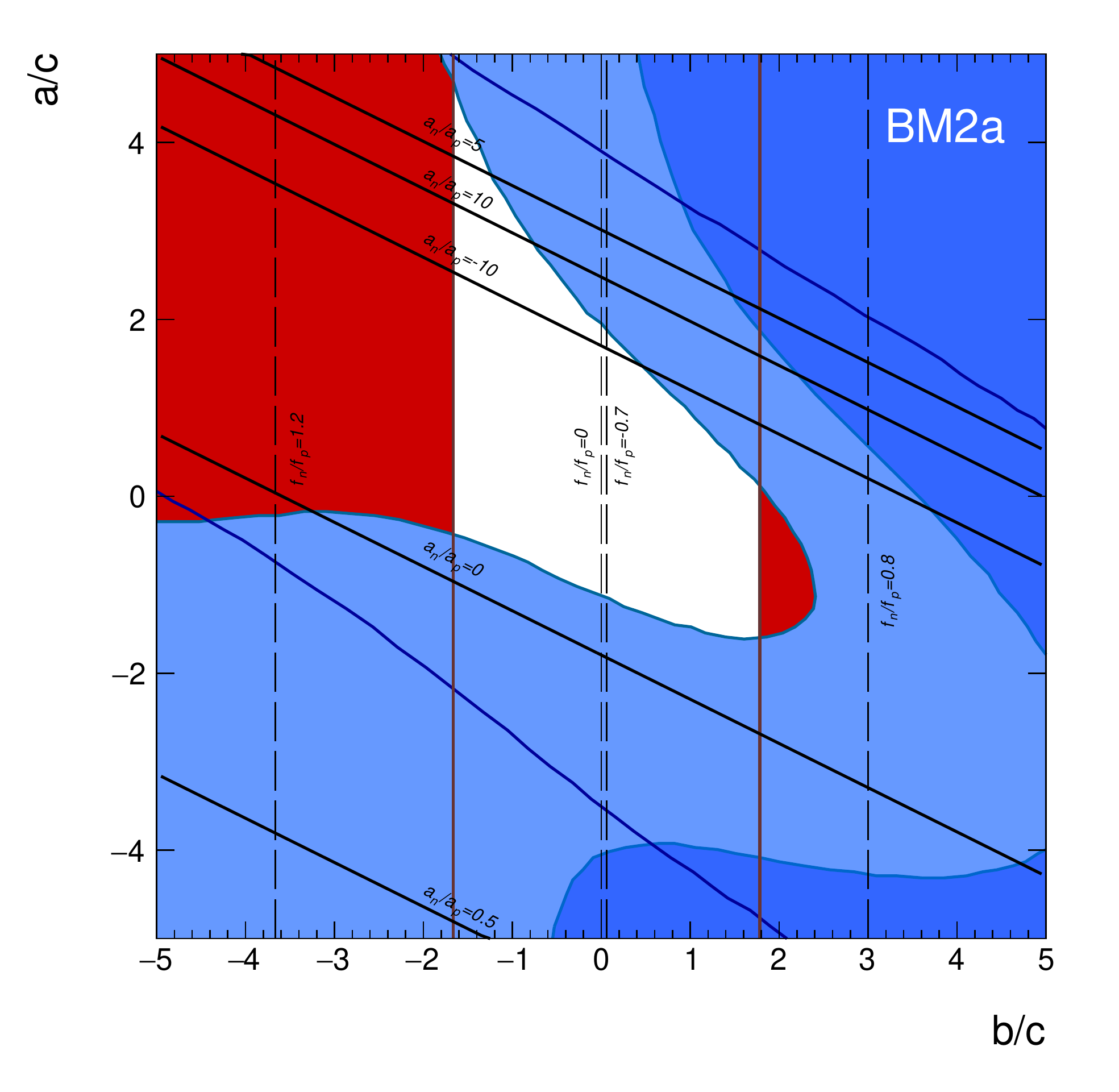}
\label{fig:BM}
\caption{$a/c$ versus $b/c$ for BM2 (left) and BM2a (right), see the text for information about the benchmark points. The LUX bound excludes all the region depicted in red. LHC bounds rule out different regions for $e^+e^-$ (light blue) and $\mu^+\mu^-$ resonances (darker blue). The dijet bound is shown also in blue and has a elliptical shape.}
\end{center}
\end{figure}
In order to compare our model with the existing data from the direct detection LUX experiment \cite{30} we calculate the number of events that it has been observed. In order to do this we include all the conditions, efficiencies and luminosity that the LUX experiment uses. In order to compute the number of events we first calculate the proton-DM cross section for spin-independent and spin-dependent scatterings,
\begin{eqnarray}
\sigma_p^{SI}&=&\frac{\mu_p^2h^2}{\pi m_{Z'}^4}(3b+c)^2,\\
\sigma_p^{SD}&=&\frac{3\mu_p^2h^2}{\pi m_{Z'}^4}[(2a+b-c)\Delta_u^p+ (2a+b+c)(\Delta^p_d + \Delta_s^p)]^2.
\end{eqnarray}
To compute the cross section for neutrons one has to multiply the SI component by a factor $(f_n/f_p)^2$ and the SD component by a factor $(a_n/a_p)^2$.

Also the $Z'$ could be produced in the LHC decaying into leptons and jets giving a distinctive signal. We have used the ATLAS detector search for high mass resonances decaying into a pair of leptons ($e^+e^-$, $\mu^+\mu^-$) for a center of mass energy $\sqrt{s}= 8$ TeV \cite{31}. In that sense we have computed the dilepton cross section using the narrow width approximation \cite{32}, 
\begin{equation}
\sigma_{l^+l^-}=\frac{\pi}{48 s} \mathcal{W}_{Z'}(s,m^2_{Z'})\times BR(Z'\rightarrow l^+l^-),
\label{eq:dilepton}
\end{equation}
where $BR(Z'\rightarrow l^+l^-)$ is the branching ratio of the $Z'$ into leptons and $\mathcal{W}_{Z'}(s,m^2_{Z'})$ is a function that contains information about the $Z'$ couplings, mass and the center-of-mass energy and provides the $Z'$ production. In the case of the dijet searches \cite{33,34} a similar approach can be done substituting the branching ratio to leptons for the one to quarks in Eq.~(\ref{eq:dilepton}).

In Fig.~(\ref{fig:BM}) we have illustrated how these constraints can exclude some regions of the parameter space. For the benchmark point of Fig.~(\ref{fig:BM}) we have chosen $c=0.1$, $d/c=3$, $h=0.5$, $m_{DM}=500$ GeV and $m_{Z'}=3$ TeV (left panel) and $c=0.05$, $d/c=5$, $h=0.25$, $m_{DM}=500$ GeV and $m_{Z'}=3$. The red area is excluded by LUX experiment. The LHC dilepton searches are depicted in blue, where $e^+e^-$ is the light blue area and $\mu^+\mu^-$ is the darker blue area. The dijet bound is shown also in blue and has a elliptical shape. As we can see the LUX bound is sensitive to the variable $b/c$ while let $a/c$ unaffected. This is because the important contribution to the direct detection searches comes from the SI cross section that only depends on the parameters $b$ and $c$. LHC searches are sensitive to the other parameters as can be seen in Fig.~(\ref{fig:BM}) for both panels. It is clear from the pictures that both direct detection and LHC searches are complementary since both impose limits in such a way that they could constraint large areas of the parameter space from this benchmark points.

\section{Conclusions}
We have studied the phenomenological features of a St\"uckelberg portal that connect two different sectors, the SM and the DM ones. These models can be embedded easily in the context of string theory, in particular in those with intersecting D6-branes. Choosing a particular set of models, called \emph{Madrid} models, that include only six phenomenologically important parameters, namely $a$, $b$, $c$, $d$, $m_DM$ and $m_{Z'}$,  we have performed a rigorous analysis of how the lightest $Z'$ boson interacts with the SM and DM sectors. We have seen that in general isospin violation arises from these interactions in contrast to other kind of portals that are common in the literature.

Furthermore we have analysed the impact of the phenomenology of this kind of portals to the recent direct detection searches as LUX and the LHC searches through the dilepton and dijet analysis performed by the ATLAS and CMS experiments. This analysis tells us that both collider and direct detection seaches are complementaries when one tries to constraint the parameter space of these models. The LUX experiment is very efficient constraining the $a/c$ parameter, however it is not sensitive to $b/c$. For that reason the use of the LHC searches are necessary since they are sensitive to all the parameters of the model.

This work shows how the nature of dark matter can be explained within the context of St\"uckerlberg portals in which the lightest $Z'$ boson can connect both SM and DM sectors. Moreover the complementarity of direct detection searches and collider searches could disentangle in a near future the properties of the $Z'$ and the dark matter.
\acknowledgments

The authors would like to thank the support of the European Union under the ERC Advanced
Grant SPLE under contract ERC-2012-ADG-20120216-320421, the support of the Consolider-Ingenio
2010 programme under grant MULTIDARK CSD2009-00064, the Spanish MICINN under Grant No.
FPA2012-34694, the Spanish MINECO ``Centro de excelencia Severo Ochoa Program'' under Grant
No.   SEV-2012-0249.

\end{document}